\documentclass[11pt]{article}
\pdfoutput=1

\usepackage{jheppub} 
\usepackage{amsmath,amssymb,epsfig,amsfonts}
\usepackage{epsf}
\usepackage{makeidx}
\usepackage{slashed}
\usepackage{verbatim} 
\usepackage{float}
\restylefloat{table}
\usepackage{pdflscape}
\usepackage{tikz}
\usepackage{pifont}
\usepackage{array}
\usepackage{youngtab}
\usepackage{multirow}
\usepackage{xcolor}
\usepackage{ulem}
\usepackage{cleveref}
\usepackage{tensor}
\usepackage{todonotes}
\usepackage{mathrsfs}
\usepackage{changepage}
\usepackage{anyfontsize}
\usepackage{caption}

\makeatletter

\newcommand{\be}{\begin{equation}}
\newcommand{\ee}{\end{equation}}
\newcommand{\ba}{\begin{aligned}}
\newcommand{\ea}{\end{aligned}}

\newcommand{\dd}{\mathrm{d}}

\newcommand{\me}{\mathrm{e}}

\newcommand{\vol}{\mathrm{vol}}

\newcommand{\ls}{\ell_s}
\newcommand{\lp}{\ell_{p}}

\makeatother

\begin{document}

\baselineskip=18pt  
\numberwithin{equation}{section}  
\allowdisplaybreaks  


%
%


\thispagestyle{empty}


\vspace*{1cm} 
\begin{center}
{\LARGE  Universal spindles: D2's on $\Sigma$ and  M5's on $\Sigma\times \mathbb{H}^3$} 
 \vspace*{1.8cm}
 
 \begin{center}

{\fontsize{12.3}{17}\selectfont Christopher Couzens$^{a ,}$\footnote{cacouzens@khu.ac.kr}, Koen Stemerdink$^{b ,}$\footnote{k.c.stemerdink@uu.nl}
}
\end{center}
\vskip .2cm

 \vspace*{0.5cm} 
 
$^{a}$ \emph{Department of Physics and Research Institute of Basic Science,\\
  Kyung Hee University, Seoul 02447, Republic of Korea}\\
 
$^{b}$\emph{ Institute for Theoretical Physics, Utrecht University \\
 Princetonplein 5, 3584 CC Utrecht, The Netherlands}\\
  
 {\tt {}}

\vspace*{0.8cm}
\end{center}

 \renewcommand{\thefootnote}{\arabic{footnote}}

\begin{adjustwidth}{0.26in}{0.26in}

\begin{center} {\bf Abstract } \end{center}

\vspace{0.32cm}

\noindent

We study the uplift of a 4d black hole in 4d Einstein--Maxwell supergravity with a spindle horizon to massive type IIA and 11d supergravity on $S^4\times \mathbb{H}^3$. The solutions exhibit features distinct to the uplift of the same 4d solution to 11d supergravity on an $S^7$. In particular, whereas the orbifold singularities may be removed in the 11d uplift on an $S^7$, they are always present in both of the classes considered here. We compute the free energy of the solutions giving predictions for the free energy of a class of 3d Chern--Simons theories wrapped on a spindle. 

\end{adjustwidth}

\noindent 



\newpage

\tableofcontents
\printindex


\section{Introduction}

The typical paradigm for studying SCFTs wrapped on compact manifolds considers branes wrapped on manifolds of constant curvature \cite{Maldacena:2000mw}. Supersymmetry of the compactified SCFT is preserved by performing a partial topological twist of the theory. Recently it has been appreciated that there are more general setups away from constant curvature manifolds that one may consider. D3-branes wrapping two-dimensional orbifolds were first discussed in \cite{Ferrero:2020laf}. Known as a spindle, the two-dimensional orbifold is isomorphic to $\mathbb{WCP}^1_{n_{\pm}}$, i.e. a sphere with conical deficits at both poles with orbifold weights $n_{\pm}$. This requires the space to have a metric of non-constant curvature. 
Since then spindle solutions have been found for other brane setups: D3-branes \cite{Ferrero:2020laf,Hosseini:2021fge,Boido:2021szx}, M2-branes \cite{Ferrero:2020twa,Cassani:2021dwa,Ferrero:2021ovq,Couzens:2021rlk,Couzens:2021cpk}, M5-branes \cite{Boido:2021szx,Ferrero:2021wvk} and the D4-D8 brane system \cite{Faedo:2021nub,Giri:2021xta}.

One interesting aspect of spindle solutions is the way in which supersymmetry is preserved. Rather than a topological twist, supersymmetry is preserved via either a \emph{topological topological twist} (also known as \emph{twist}) or an \emph{anti-twist}, two new mechanisms of preserving supersymmetry \cite{Ferrero:2021etw}. In both cases the R-symmetry is mixed with the azimuthal rotations of the spindle, and the two twists can be differentiated by the amount of R-symmetry flux threading through the spindle. As in the regular topological twist, the amount of flux threading through the spindle equals the Euler characteristic of the spindle for a (topological topological) twist. However, in contrast to the the usual topological twist, the local curvature of the R-symmetry gauge field is not equal to the curvature of the tangent bundle. The anti-twist is even further away from the familiar topological twist: here the total R-symmetry flux threading through the spindle is not even equal to the Euler characteristic. To be more concrete, let us denote the total R-symmetry flux threading through the spindle by $Q_R$ and the Euler-characteristic to be $\chi(\Sigma)$. Then for such spindle solutions we have
\be
Q_R=\frac{n_-+\sigma n_+}{n_-n_+}\, ,\qquad \chi(\Sigma)= \frac{n_-+n_+}{n_- n_+}\, ,
\ee
with $\sigma=1$ for the twist and $\sigma=-1$ for the anti-twist.

Concurrently to the spindle story, M5-branes wrapped on another 2d space of non-constant curvature were also studied \cite{Bah:2021hei,Bah:2021mzw}. These spaces are topologically discs, admitting a conical deficit at the centre of the disc and a boundary along which the solution is singular. This singularity is related to the presence of smeared M5-branes and the disc can be identified as a sphere with two punctures; one a maximal regular puncture giving a conical defect, as in the spindle class of solutions, and the other an irregular puncture leading to the singular boundary. It was shown in \cite{Bah:2021hei,Bah:2021mzw} that these solutions are dual to a class of Argyres--Douglas theories. The solution can be further generalised to replace the regular puncture of maximal type, to a regular puncture labelled by an arbitrary Young diagram giving a partition of a flux number, \cite{Couzens:2022yjl}.\footnote{It would be interesting to understand the field theory dual of the M5-brane on a spindle to a similar level of M5-branes where the higher amount of supersymmetry allows for more control on the field theory side. }

Disc solutions have also been found for other brane configurations since the first discs: D3-branes in \cite{Couzens:2021rlk, Suh:2021ifj}, D4-D8 branes in \cite{Suh:2021aik}, M2-branes \cite{Couzens:2021rlk,Suh:2021hef}, M5-branes \cite{Bah:2021hei,Bah:2021mzw,Karndumri:2022wpu,Couzens:2022yjl}. As shown in \cite{Couzens:2021rlk,Couzens:2021tnv}, for D3 and M2-branes respectively, though the logic applies more widely, discs are different global completions of the local solutions giving rise to multi-charge spindle solutions and preserve twice the supersymmetry as the corresponding spindle solution.\footnote{One of the disc solutions in \cite{Karndumri:2022wpu} is somewhat of an outlier here as it preserves $\mathcal{N}=1$ only.} A third, and final interesting related solution are the defect solutions of \cite{Gutperle:2022pgw}. There M5-branes wrap a non-compact two-dimensional base giving rise to a co-dimension 2 defect in the 6d world-volume theory on a stack of M5-branes. As for the discs, this is a different global completion of the same local solution giving rise to the multi-charge spindle solutions.\footnote{Similar defect solutions are expected for all the other brane configurations above where spindle solutions have been found. }
A recent further extension is to consider branes wrapping 4d orbifold geometries, see for example \cite{Cheung:2022ilc} and \cite{Suh:2022olh} for the product of a spindle with a Riemann surface.

Here we consider two classes of solutions obtained by uplifting the same 4d solution of 4d $\mathcal{N}=2$ Einstein--Maxwell supergravity. The first is rotating D2-branes in massive type IIA string theory wrapped on a spindle, which uses the truncation of  massive type IIA to 4d $\mathcal{N}=2$ Einstein--Maxwell supergravity \cite{Varela:2019vyd}. Our second class exploits a second consistent truncation to 4d $\mathcal{N}=2$ Einstein--Maxwell supergravity on a class of geometries describing M5-branes wrapped on a SLAG 3-manifold \cite{Gauntlett:2007ma}. 
In order to keep this paper as self-contained as possible we review the 4d solution in detail, in particular the conical deficits at the poles and the type of twist required to preserve supersymmetry. 

Upon uplifting the solution to massive type IIA we find that the internal 8d geometry of the 10d solution takes the form of a SE$_5\times \text{S}^1_{z}$ fibration over a rectangle, while the uplift to the SLAG geometry gives rise to a $S^2\times S^1\times S^1\times \mathbb{H}^3$ fibration over a rectangle. In contrast to the M2 brane uplift of the same solution, we find that the internal spaces \emph{cannot} be made smooth, instead the conical deficits remain in the uplifted solution. This behaviour is more reminiscent of M5-branes on a spindle \cite{Ferrero:2021wvk}, yet the way in which supersymmetry is preserved is different. In the D2-brane case considered here supersymmetry is preserved with an anti-twist, while the M5-brane case uses a (topological topological) twist. Consequently, the presence of (monopole) singularities in the uplift not a property exclusive to either of the twists but rather a property of the manner in which one uplifts the solution as one may somewhat surmise. 

We begin by reviewing the rotating spindle solution in 4d $\mathcal{N}=2$ Einstein--Maxwell supergravity in section \ref{eq:EMspindle} before studying its uplift to massive type IIA supergravity in section \ref{sec:IIA}. In section \ref{sec:SLAG} we uplift the solution to 11d supergravity on $S^4\times \mathbb{H}^3$ giving rise to the holographic duals of M5-branes wrapped on the 5d product space $\Sigma\times \mathbb{H}^3$. We conclude in \ref{Sec:Conclude} and relegate a review of the regularity of the M5-brane on a spindle solution to an appendix \ref{app:M5}.

\section{Spindle in 4d Einstein--Maxwell supergravity}\label{eq:EMspindle}

In this section we will consider the spindle solution of 4d $\mathcal{N}=2$ Einstein-Maxwell supergravity. This was first discussed in the context of a spindle geometry in \cite{Ferrero:2020twa}, having been originally found in \cite{Plebanski:1976gy}, and we will briefly review the analysis of the former paper. First, we discuss the full black hole solution, and afterward we discuss the near-horizon limit.

\subsection{Full solution}

The action for 4d $\mathcal{N}=2$ Einstein--Maxwell supergravity we will use is
\be
S=\frac{1}{16 \pi G_{(4)}}\int \Big((R+ 6 )\star 1 -F\wedge \star F\Big)\, ,
\ee
where we have normalised the cosmological constant for simplicity. A solution to this action is\footnote{We follow the presentation of the solution in \cite{Ferrero:2020twa} for ease of comparison with existing literature.}
\begin{align}
\begin{aligned}
\dd s^2&=\frac{1}{H(r,\theta)^2}\bigg[-\frac{Q(r)}{S(r,\theta)}\Big(\dd t -a \sin^2\theta\, \dd\phi\Big)^2+\frac{S(r,\theta)}{Q(r)}\,\dd r^2 \\
&\quad\; +\frac{P(\theta)}{S(r,\theta)}\,\sin^2\theta\, \Big(a\,\dd t-(r^2+a^2)\,\dd\phi\Big)^2 +\frac{S(r,\theta)}{P(\theta)}\,\dd\theta^2\bigg]\, ,\\[4pt]
A&=-\frac{e r}{S(r,\theta)}\big(\dd t -a \sin^2\theta \dd\phi\big)+\frac{g \cos \theta}{S(r,\theta)}\big(a\dd t -(r^2+a^2)\dd\phi\big)\, ,
\end{aligned}
\end{align}
where
\begin{align}
\begin{aligned}
H(r,\theta)&=1-\alpha\, r \cos \theta\, ,\\
S(r,\theta)&=r^2+a^2 \cos^2\theta\, ,\\
P(\theta)&=1-2\alpha m \cos\theta +\Big(\alpha^2 (a^2+e^2+g^2)-a^2\Big)\cos^2\theta\, ,\\
Q(r)&=\Big(r^2-2 mr +a^2 +e^2+ g^2\Big)(1-\alpha^2 r^2)+(a^2+r^2) r^2\,.
\end{aligned}
\end{align}
We demand that the metric has the correct signature. This requires that the functions $S(r,\theta)$, $Q(r)$ and $P(\theta)$ are all positive. The range of $\theta$ is taken to be $[0,\pi]$ and it is convenient to define $\theta_-=0,\, \theta_+=\pi$. The requirement of the correct metric signature then implies
\be
\begin{split}
P(\theta_-)&=1-2\alpha m +\Big(\alpha^2 (a^2+e^2+g^2)-a^2\Big)>0 \, ,\\
P(\theta_+)&=1+2\alpha m +\Big(\alpha^2 (a^2+e^2+g^2)-a^2\Big)>0 \,.
\end{split}
\ee
Following \cite{Ferrero:2020twa} we study the metric on constant $(t,r)$ slices, expanding around $\theta_{\pm}$ the $\theta,\phi$ terms of the metric become
\be\label{metric spindle general}
\dd s^2\sim \frac{1}{H(r,\theta_{\pm})^2}\frac{r^2+a^2}{P(\theta_{\pm})}\Big[\dd\theta^2+(\theta-\theta_\pm)^2P(\theta_{\pm})^2 \,\dd\phi^2\Big]\, .
\ee
If the geometry is regular round two-sphere one requires that $P(\theta_{+})^2=P(\theta_-)^2$, however one finds
\be
P(\theta_+)-P(\theta_-)=4\alpha m\, ,
\ee
and therefore it is not possible to assign a period to $\phi$ such that \eqref{metric spindle general} is smooth at both poles when $\alpha\neq 0$, i.e. it cannot be a smooth $S^2$. Instead one must contend with conical singularities at the poles. The period of $\phi$ is then taken to be
\begin{equation}
\Delta \phi=\frac{2\pi}{P(\theta_+)\,n_+}=\frac{2\pi}{P(\theta_-)\,n_-}\, ,
\end{equation}
with $n_{\pm}$ positive relatively prime integers. This then defines a spindle which is the weighted projective space $\mathbb{WCP}^{1}_{[n_-,n_+]}$.

\subsection{Near-horizon limit}

\begin{figure}[h!]
\centering
  \includegraphics[width=0.75\linewidth]{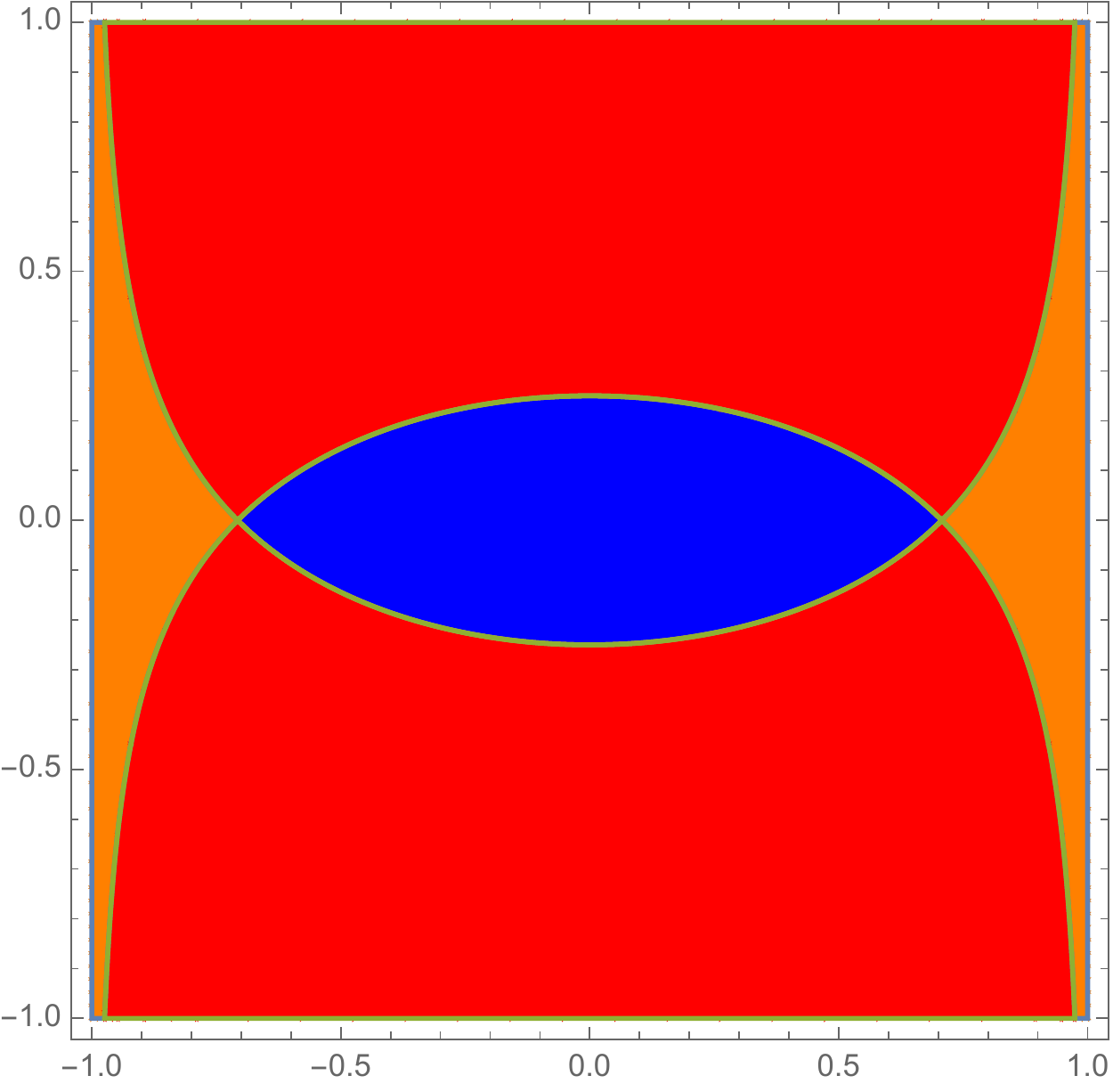}
  \captionsetup{width=.85\linewidth}
  \caption{\textit{The domain of validity for the roots of the quartic $f(w)$ in $j$ (horizontal axis) and $q$ (vertical axis) parameter space. The central blue region has four real roots, the left and right orange regions have no real roots and the top and bottom red regions have two real roots. The boundaries of the regions are where there are two equal roots. The boundary from the upper half is where $w_3=w_4$ and the boundary form the bottom half is where $w_1=w_2$. }}
  \label{fig:rootdomain}
\end{figure}

If we impose extremality and take the near-horizon limit of the black hole solution presented earlier in this section, we obtain a fibered AdS$_2$ space, see \cite{Ferrero:2020twa}. The near-horizon geometry is given by
\begin{align}
\begin{aligned}
\dd s^2_\text{NH}&=\sqrt{P(w)}\,\bigg[-r^2\,\dd t^2+\frac{\dd r^2}{r^2}+\dd s^2_\Sigma\bigg]\, ,\\[4pt]
\dd s^2_\Sigma &= \frac{1}{f(w)}\,\dd w^2+\frac{f(w)}{P(w)}\,(\dd z+j r \,\dd t)^2\,,\\[6pt]
A&=h(w)\big(\dd z+j r\, \dd t\big)\, ,
\end{aligned}
\end{align}
with\footnote{We have performed some redefinitions of the functions and coordinates in comparison with \cite{Ferrero:2020twa} since it puts the the functions into a more succinct form. }
\begin{align}
\begin{aligned}
P(w)&=\bigg[(w-q)^2 +\frac{j^2}{4}\bigg]^2 \, ,\\[2pt]
f(w)&= P(w)-\big(1-j^2\big)w^2\, , \\[2pt]
h(w)&=-\sqrt{1-j^2}\bigg[ \frac{4 (w-q)(w+q)-j^2 }{4 (w-q)^2+j^2}\bigg]\, .
\end{aligned}
\end{align}
In order to have a well-defined solution of correct signature, we need the coordinate $w$ to be defined on an interval where both $f(w)$ and $P(w)$ are non-negative. Note that $P(w)$ is strictly positive since it is a square, and therefore the correct signature requires an interval between which $f(w)$ is positive and vanishes at the end-points.
This implies that $f(w)$ must have four real roots, and the interval then lies between the two middle roots. We denote the four roots by $w_1,\ldots,w_4$, where we assign the labels in ascending order. One can readily compute expressions for the roots, and it can be checked that there is only one labelling consistent with $w_1\leq w_2<w_3\leq w_4$ (we don't allow for $w_2=w_3$ as this would not give a finite interval for $w$). We find the roots, labeled in ascending order, to be
\begin{equation}
\begin{alignedat}{2}
w_1 &= \frac{1}{2}\bigg(2q -\sqrt{1-j^2}-\sqrt{1-2j^2-4q\sqrt{1-j^2}}\bigg) \,,\\
w_2 &= \frac{1}{2}\bigg(2q -\sqrt{1-j^2}+\sqrt{1-2j^2-4q\sqrt{1-j^2}}\bigg) \,,\\
w_3 &= \frac{1}{2}\bigg(2q +\sqrt{1-j^2}-\sqrt{1-2j^2+4q\sqrt{1-j^2}}\bigg) \,,\\
w_4 &= \frac{1}{2}\bigg(2q +\sqrt{1-j^2}+\sqrt{1-2j^2+4q\sqrt{1-j^2}}\bigg) \,.
\end{alignedat}
\end{equation}
For convenience it is useful to redefine $w_2$ and $w_3$ to be $w_-$ and $w_+$ respectively, and the interval for $w$ is then $w\in [w_-,w_+]$. Furthermore, it can be shown that zero must always lie in this interval, i.e. $w_-$ and $w_+$ have opposite sign. For all four of the roots to be real we must impose 
\be
-\frac{1}{\sqrt{2}}\leq j\leq \frac{1}{\sqrt{2}}\, ,\quad \frac{2 j^2-1}{4 \sqrt{1-j^2}}\leq q\leq \frac{1-2j^2}{4 \sqrt{1-j^2}}\, .
\ee
In figure \ref{fig:rootdomain} we have plotted the parameter space of the roots. The blue region is the region with four real roots, the red is the region with two real roots and the orange is the region with no real roots. 
The boundaries between the different locations are the regions with double roots, note that there are two special points in parameter space at $j=\pm 2^{-1/2}, q=0$ where there are two pairs of double roots. In figure \ref{fig:rootdomain} these points are located where the three different colours meet. The double roots leads to a degeneration of the form of a (quotient of) 2d hyperbolic space, $\mathbb{H}^2$ rather than the more canonical $\mathbb{R}^2$ quotient studied here. In fact these solutions are the ones considered in \cite{Klemm:2014rda}, see also \cite{Gnecchi:2013mja}, where one replaces the conical defects of the spindle with cusps. Note that two cusps are not possible for a static geometry, in that case only one cusp is possible leading to a black bottle geometry.

Just as for the full black hole solution, we find that the geometry cannot be made smooth at both poles, i.e. $\Sigma$ is a spindle. The period of $z$ is 
\be
\frac{\Delta z}{2\pi}=\frac{2\sqrt{P(w_{\pm})}}{|f'(w_{\pm})|\,n_{\pm}}\, ,\label{eq:Delz}
\ee
with $n_{\pm}$ positive relatively prime integers. 
For later it is sometimes convenient to define the $2\pi$-periodic coordinate $\hat{z}$ via,
\be
\hat{z}=\frac{2\pi}{\Delta z}   \, {z} \, .\label{eq:zhat}
\ee

In order to determine via which mechanism supersymmetry is preserved we compute both the magnetic charge and the Euler characteristic of the spindle, and compare them. First, the total magnetic charge is computed as an integral of the field strength
\begin{equation}
Q_m= \frac{1}{2\pi}\int_\Sigma \dd A = -\frac{\Delta z}{2\pi}\frac{8\sqrt{1-j^2}\;w(w-q)}{j^2+4(w-q)^2}\,\bigg|^{w=w_+}_{w=w_-} = -\frac{\Delta z}{2\pi}\frac{\sqrt{1-j^2}\;w\,P'(w)}{2P(w)}\,\bigg|^{w=w_+}_{w=w_-} \,.
\end{equation}
Since $w_\pm$ are roots of $f(w)$, it follows that $P(w_\pm) = (1-j^2)w_\pm^2$ and further that the first derivative satisfies $P'(w_\pm) = f'(w_\pm)+2(1-j^2)w_\pm$. Using these identities, we find the expression
\begin{equation}\label{mag charge}
Q_m = \frac{\Delta z}{4\pi\sqrt{1-j^2}}\bigg( \text{sgn}(w_+)\frac{|f'(w_+)|}{|w_+|} + \text{sgn}(w_-)\frac{|f'(w_-)|}{|w_-|} \bigg) \,,
\end{equation}
for the magnetic charge of the near-horizon solution. The electric charge is defined by
\be
Q_e=\frac{1}{2\pi}\int_{\Sigma} \star  F\, ,
\ee
which gives
\be
Q_e=\frac{\Delta z}{2\pi} \frac{ j \sqrt{1-j^2} w}{\sqrt{P(w)}}\bigg|_{w=w_-}^{w=w_+}=j\frac{\Delta z}{\pi}\, . \label{eq:electric}
\ee

The Euler characteristic can be computed as an integral of the 2d Ricci scalar over the Riemann surface. We find
\begin{equation}
\chi(\Sigma) = \frac{1}{4\pi}\int_\Sigma R_2\: \text{dvol}_\Sigma = \frac{f(w)P'(w)-P(w)f'(w)}{P(w)^{3/2}}\,\frac{\Delta z}{4\pi}\,\bigg|^{w=w_+}_{w=w_-} \,,
\end{equation}
which, after using the properties of $f(w)$ and $P(w)$ evaluated at the roots of $f(w)$, yields the expression
\begin{equation}\label{euler}
\chi(\Sigma) = \frac{\Delta z}{4\pi\sqrt{1-j^2}}\bigg( \frac{|f'(w_+)|}{|w_+|} + \frac{|f'(w_-)|}{|w_-|} \bigg) \,.
\end{equation}
Application of the expression for the period in \eqref{eq:Delz} gives
\be
Q_m=\frac{\text{sgn}(w_+)}{n_+}+\frac{\text{sgn}(w_-)}{n_-}\, ,\qquad \chi(\Sigma)= \frac{1}{n_+}+\frac{1}{n_-}\, .
\ee
We see that the charge \eqref{mag charge} and the Euler characteristic \eqref{euler} are equal if $w_-$ and $w_+$ have the same sign, and unequal if they have opposite signs. We found earlier in this section that zero always lies in the interval $[w_-,w_+]$, so we always find the latter case. This implies that supersymmetry is preserved with a mechanism known as an anti-twist, rather than with a twist, following the language of \cite{Ferrero:2021etw,Couzens:2021cpk}.

Before we close this section let us give the roots in terms of the orbifold weights. To solve the period condition \eqref{eq:Delz} 
we must impose
\begin{align}
q=\frac{(1-2j^2)(n_-^2-n_+)^2}{4 \sqrt{1-j^2}(n_-^2+n_+^2)}\, .
\end{align}
We may further eliminate $j$ in favour of the electric charge by using \eqref{eq:electric},
\be
j=\frac{n_- n_+ Q_e}{\sqrt{2(n_-^2+n_+^2+n_-^2n_+^2 Q_e^2)}}\, ,
\ee 
and the roots become
\begin{align}
w_+&=\frac{n_+^2 +3n_-^2+ n_-^2n_+^2 Q_e^2-2 n_-\sqrt{2 (n_-^2+n_+^2 )+n_-^2n_+^2 Q_e^2}}{2 \sqrt{2}\sqrt{\big(n_-^2+n_+^2+n_-^2n_+^2 Q_e^2\big)\big(2(n_-^2+n_+^2)+n_-^2 n_+^2 Q_e^2\big)}}\, ,\\
w_-&=\frac{-(3n_+^2 +n_-^2+ n_-^2n_+^2 Q_e^2)+2 n_+\sqrt{2 (n_-^2+n_+^2 )+n_-^2n_+^2 Q_e^2}}{2 \sqrt{2}\sqrt{\big(n_-^2+n_+^2+n_-^2n_+^2 Q_e^2\big)\big(2(n_-^2+n_+^2)+n_-^2 n_+^2 Q_e^2\big)}}\, .
\end{align}

\section{Uplift to massive type IIA}\label{sec:IIA}

In the previous section we have reviewed the rotating spindle that can be constructed in 4d $\mathcal{N}=2$ Einstein--Maxwell supergravity. We will now consider uplifting the solution to massive type IIA supergravity focussing on the uplift giving rise to the holographic duals of 3d Chern--Simons SCFTs arising from D2 branes wrapped on a spindle. Massive type IIA supergravity on a topological six-sphere gives rise to 4d $\mathcal{N}=8$ ISO(7) supergravity. As shown in \cite{Varela:2019vyd} there exists a consistent truncation of this theory to 4d $\mathcal{N}=2$ Einstein--Maxwell supergravity. We may therefore uplift the known solution reviewed in the previous section to massive type IIA and study its properties using the uplifting formula presented in \cite{Varela:2019vyd}. In general the uplift of a solution of 4d $\mathcal{N}=2$ Einstein--Maxwell in Einstein frame is
\begin{equation}\label{10d metric}
\begin{aligned}
\dd s^2_{10} &= L^2\,(2+\cos^2\alpha)^{1/8} \sqrt{1+\cos^2\alpha}\:\bigg[\frac{1}{3}\dd s^2_4+\dd s^2_{Y_6}\bigg] \,, \\[4pt]
\dd s^2_{Y_6}&=\frac{1}{2}\,\dd \alpha^2 +\frac{3 \sin^2\alpha}{4+2\cos^2\alpha}\,\hat{\eta}^2+\frac{\sin^2\alpha}{1+\cos^2\alpha}\,\dd s^2_{\text{KE}_4}\, ,
\end{aligned}
\end{equation}
where 
\begin{align}
\hat{\eta}&=\eta +\frac{1}{3}A=\dd\psi+\sigma+\frac{1}{3}A\, ,\\
\dd\eta&=2\, J_{\text{KE}_4}\, ,
\end{align}
and $A$ is the 4d gauge field. The coordinate $\alpha$ is defined on an interval $[0,\pi]$, and the metric on $Y_6$ is that of a sine cone, which is squashed due to the $\alpha$-dependent prefactors. The K\"ahler--Einstein metric is normalized to satisfy $R_{mn}=6 g_{mn}$ and $J_{\text{KE}_4}$ is the K\"ahler form. Together $\eta$ and $\dd s^2_{\text{KE}_4}$ form a squashed 5d Sasaki--Einstein manifold.

The other fields in the uplift to massive type IIA are given by
\begin{align}
\me^{\Phi}&=\frac{1}{m^{4/5}}\frac{(2+\cos^2\alpha)^{3/4}}{1+\cos^2\alpha}\, ,\label{eq:dil}\\
H_3&=\frac{L^{2}}{m^{2/5}}\bigg[\frac{2 \sin^3 \alpha}{(1+\cos^2\alpha)^2}\,J_{\text{KE}_4}\wedge \dd \alpha +\frac{1}{2\sqrt{3}}\sin\alpha\, \dd \alpha \wedge\star_4 F\bigg]\, ,\label{eq:H}\\
F_0&=\frac{m}{L}\, ,\label{eq:F0}\\
F_{2}&=L \,m^{3/5}\bigg[ -\frac{\sin^2\alpha\, \cos\alpha}{(2+\cos^2 \alpha)(1+\cos^2\alpha)}\,J_{\text{KE}_4}-\frac{3\sin\alpha\,(2-\cos^2 \alpha)}{2(2+\cos^2 \alpha)^2}\,\dd \alpha \wedge \hat{\eta}\nonumber\\
&\;\;\;\;+\frac{\cos\alpha}{2(2+\cos^2\alpha)}\,F-\frac{1}{2\sqrt{3}}\cos\alpha \,\star_4 \!F\bigg]\, ,\label{eq:F2}\\
F_{4}&=L^{3}\,m^{1/5}\bigg[\frac{\sin^4\alpha\,(2+3 \cos^2\alpha)}{2(1+\cos^2\alpha)^2}\,J_{\text{KE}_4}^2 +\frac{3\sin^3\alpha\, \cos\alpha\,(4+\cos^2\alpha)}{2(1+\cos^2\alpha)(2+\cos^2 \alpha)}\,J_{\text{KE}_4}\wedge \dd \alpha \wedge \hat{\eta}\nonumber\\
&\;\;\;\;+\frac{1}{\sqrt{3}}\,\dd\vol_4-\frac{1}{4}\sin \alpha\,\cos\alpha\,\Big(\frac{2\sin \alpha\,\cos\alpha}{1+\cos^2\alpha}\,J_{\text{KE}_4}+\dd\alpha\wedge \hat{\eta}\Big)\wedge F \nonumber\\
&\;\;\;\;-\frac{1}{4\sqrt{3}}\Big( \frac{2\sin^2\alpha}{1+\cos^2\alpha}\,J_{\text{KE}_4}+\frac{3\sin\alpha\,\cos\alpha}{2+\cos^2\alpha}\,\dd\alpha\wedge \hat{\eta} \Big)\wedge\star_4 F\bigg]\, .\label{eq:F4}
\end{align}
We have redefined $L$ and $m$ with respect to some of the existing literature in order to make $L$ have length scaling dimension 1 and for $m$ to scaling dimension 0, where necessary we will provide the dictionary.

\subsection{Metric regularity}\label{sec:D2reg}

We now want to study the regularity of the uplifted 10d solution. The method for analysing the regularity closely follows \cite{Ferrero:2020twa} however we will see that the global structure here is different to that considered there. In particular whereas the conical singularities of the spindle may be de-singularised in the uplift of \cite{Ferrero:2020twa} it is not possible here and mimics more the behaviour of the M5-brane on a spindle solution \cite{Ferrero:2021wvk} and the D4-brane on a spindle solution \cite{Faedo:2021nub}. 

\begin{figure}[h!]
\centering
  \includegraphics[width=0.75\linewidth]{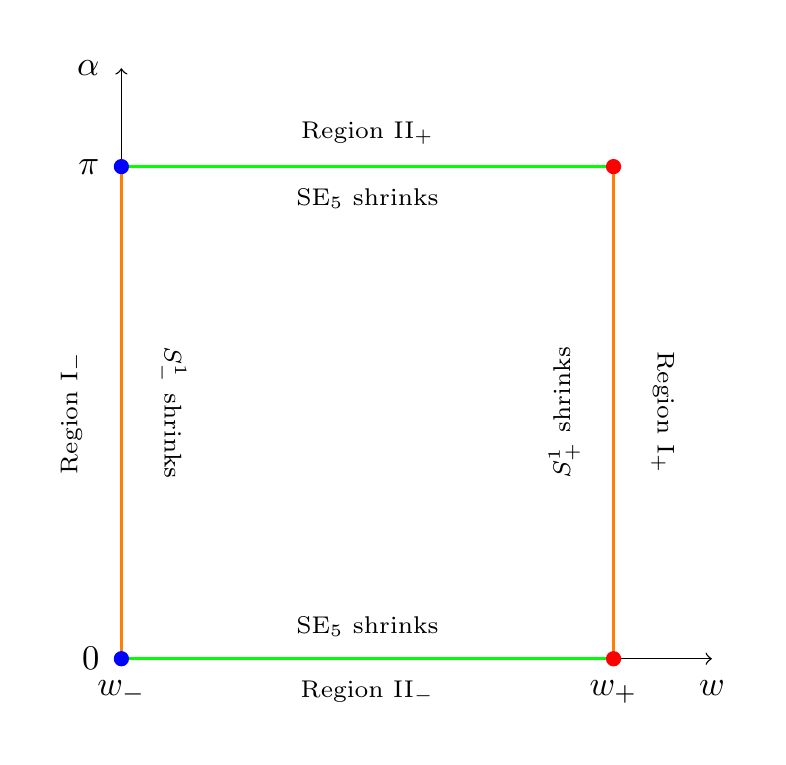}
  \captionsetup{width=.85\linewidth}
  \caption{\textit{A schematic plot of the rectangle over which there is a} SE$_5\times S^1_{z}$\textit{ fibration. A monopole is located at each corner of the rectangle and they are called region III in the main text. The red monopoles have weight $n_{+}$ whilst the blue monopoles have weight $n_-$.}}
  \label{fig:D2rectangle}
\end{figure}

We write the 10d geometry \eqref{10d metric} as
\begin{equation}
\dd s^2_{10} = L^2\,(2+\cos^2\alpha)^{1/8} \sqrt{1+\cos^2\alpha}\:\bigg[\frac{\sqrt{P(w)}}{3}\,\dd s^2_{\text{AdS}_2}+\dd s^2_8 \bigg] \,,
\end{equation}
and study the regularity of the 8d metric at fixed $(t,r)$,
\begin{equation}\label{8d metric}
\dd s^2_8 = \frac{\sqrt{P(w)}}{3f(w)}\,\dd w^2 + \frac{f(w)}{3\sqrt{P(w)}} \,D z^2 + \frac{1}{2}\,\dd \alpha^2 +\frac{3 \sin^2\alpha}{4+2\cos^2\alpha}\,\hat{\eta}^2+\frac{\sin^2\alpha}{1+\cos^2\alpha}\,\dd s^2_{\text{KE}_4} \,,
\end{equation}
where 
\be
Dz=\dd z+j r \dd t\, .
\ee
We find this 8d space to be an $S^1_z\times\text{SE}_5$ fibration over a rectangle parametrized by the coordinates $w$ and $\alpha$. A plot of this rectangle is given in figure \ref{fig:D2rectangle} where we indicate the regions where we will study regularity in detail. In the interior of the rectangle regularity is ensured, since none of the coefficients in \eqref{8d metric} vanish there.

\paragraph{Region I: the poles of the spindle.}

We will first focus on the metric away from $\alpha=0,\pi$ and study the degeneration at the two poles of the spindle. To that end we construct the following U$(1)$ Killing vectors, each of which has vanishing norm at one of the poles:
\begin{equation}
\partial_{\phi_{\pm}}=c_\pm\left(\partial_z-\frac{1}{3} h(w_{\pm})\,\partial_{\psi}\right)\, ,
\end{equation}
with $c_\pm$ constants that we will fix momentarily. By expanding the norm of these Killing vectors around $w_\pm$ we find that\footnote{Here we use the leading order expansions $f(w)= |f'(w_\pm)| \,\Delta w$ and $P(w)=(1-j^2)w_\pm^2$ around $\Delta w = 0$.}
\begin{equation}
||\partial_{\phi_{\pm}}||^2 = \frac{c_\pm^2 \,|f'(w_\pm)|}{3\sqrt{1-j^2}\,|w_\pm|} \, \Delta w + \mathcal{O}(\Delta w^2) \,,
\end{equation}
where $\Delta w = \mp(w-w_\pm)$ is the positive distance away from the pole. Consequently, the metric spanned by $w$ and $\phi_\pm$ close to $w_\pm$ can be written as
\begin{equation}
\frac{\sqrt{1-j^2}\,|w_\pm|}{3|f'(w_\pm)|\Delta w}\,\dd w^2 + \frac{c_\pm^2 \,|f'(w_\pm)|\Delta w}{3\sqrt{1-j^2}\,|w_\pm|}\,\dd \phi_\pm^2 = \frac{4\sqrt{1-j^2}\,|w_\pm|}{3|f'(w_\pm)|}\left[ \dd R^2 + \frac{c_\pm^2 \,f'(w_\pm)^2}{4(1-j^2)w_\pm^2}\, R^2 \dd \phi_\pm^2 \right]\,,
\end{equation}
where we have introduced the radial coordinate as $R^2=\Delta w$. In order for the Killing vector $\partial_{\phi_{\pm}}$ to degenerate smoothly at the pole, we see that we need to normalize it as
\begin{equation}
c_\pm = \frac{2\sqrt{1-j^2}\,w_\pm}{f'(w_\pm)}\,.
\end{equation}
The coordinates $\phi_\pm$ now have periodicity $2\pi$ around $w=w_\pm$ respectively and the conical singularities are resolved. This works in an analogous way to the discussion in \cite{Ferrero:2020twa} for the uplift of this solution to 11d supergravity on a Sasaki--Einstein 7-manifold. 

Let us study the metric close to the poles in a bit more detail. We switch to the coordinates $\phi_\pm$ using the transformations
\begin{equation}
\begin{aligned}
z &= 2\sqrt{1-j^2}\left(\frac{w_+}{f'(w_+)}\,\phi_+ - \frac{w_-}{f'(w_-)}\,\phi_- \right)\,,\\
\psi &= -\tfrac{2}{3}\sqrt{1-j^2}\left(\frac{w_+\,h(w_+)}{f'(w_+)}\,\phi_+ - \frac{w_-\,h(w_-)}{f'(w_-)}\,\phi_- \right)\,.
\end{aligned}
\end{equation}
We may rewrite the metric in these new coordinates as
\begin{align}
\dd s^2_8=\frac{\sqrt{P(w)}}{3f(w)}\,\dd w^2+\frac{1}{2}\,\dd\alpha^2 +F_{+}\,D \phi_+^2+F_-\,D\phi_-^2+G\, D\phi_+D\phi_-+\frac{\sin^2\alpha}{1+\cos^2\alpha}\,\dd s^2_{\text{KE}_4}\,,\label{eq:D2S3}
\end{align}
where 
\begin{align}
D\phi_{\pm}=\dd \phi_{\pm} + \frac{3 f'(w_\pm)}{2\sqrt{1-j^2}\,w_{\pm}(h(w_\mp)-h(w_\pm)}\left(\sigma+\tfrac{1}{3}j r \,h(w_\mp)\,\dd t\right)\,,
\end{align}
and
\begin{align}
F_{\pm}&=\frac{2(1-j^2)w_\pm^2\Big(2(2+\cos^2 \alpha)f(w)+\sin^2\alpha\big(h(w)-h(w_\pm)\big)^2\sqrt{P(w)}\Big)}{3(2+\cos^2\alpha)\sqrt{P(w)}f'(w_\pm)^2}\,,\\
G&=\frac{4(1-j^2) w_+w_-\Big(2(2+\cos^2\alpha)f(w)+\sin^2\alpha\big(h(w)-h(w_+)\big)\big(h(w)-h(w_-)\big)\sqrt{P(w)})\Big)}{3(2+\cos^2\alpha)\sqrt{P(w)}f'(w_+)f'(w_-)}\, .\nonumber
\end{align}
Note that $F_{+}(w_+)=G(w_+)=0$ and $F_{-}(w_-)=G(w_-)=0$. The $w, \phi_{\pm}$ coordinates are topologically an $S^3$ at fixed $\alpha$ and we see that we have an $S^3$ fibration over a K\"ahler--Einstein base. This is similar to the M2-brane case.

\paragraph{Region II.}

In order to study the metric close to $\alpha=0,\pi$, we simply Taylor expand the trigonometric functions in the metric around these points. This yields
\begin{equation}
\dd s^2_8 = \frac{\sqrt{P(w)}}{3f(w)}\,\dd w^2 + \frac{f(w)}{3\sqrt{P(w)}} \,D z^2 + \frac{1}{2}\Big(\dd R^2 +R^2\,\big(\hat{\eta}^2+\dd s^2_{\text{KE}_4}\big)\Big) \,,
\end{equation}
where we use $R=\alpha,\pi-\alpha$ for the distance away from $\alpha=0,\pi$ respectively. We see that this is the metric of a 6d cone fibered over the spindle. When KE$_4=\mathbb{CP}^2$ this is $\mathbb{R}^6$ and free of singularities, while for other K\"ahler--Einstein manifolds it has a singularity at the tip of the Calabi--Yau cone. This singularity appears also in the AdS$_4$ vacuum solution and is thus expected to remain.

\paragraph{Region III: monopoles.}

In addition to the limits to the endpoints of the $w$ and $\alpha$ intervals, we should also take the limit where both $w$ and $\alpha$ go to an endpoint simultaneously, i.e. to a corner of the rectangle in figure \ref{fig:D2rectangle}.

We denote the endpoints of the intervals by $\alpha\in[\alpha_{-},\alpha_+]=[0,\pi]$ and $w\in[w_-,w_+]$. In order to take the combined limit, we need a coordinate transformation to `polar coordinates' around the corner of the rectangle. A proper set of such coordinates is given by $(R,\zeta)$ defined by
\be\label{coordinate transformation Rzeta}
\alpha=\alpha_{\pm}\mp R\cos\zeta\, ,\qquad\quad w=w_{\pm}\mp \left|\frac{3 f'(w_{\pm})}{8\sqrt{1-j^2}\, {w_\pm}}\right| R^2\sin^2\zeta\, ,
\ee
where the signs are chosen such that $R$ is a positive radial coordinate for each of the four corners. The scalings and powers in \eqref{coordinate transformation Rzeta} are chosen such that we find the appropriate metric ($\dd R^2+R^2\dd\zeta^2$) in polar coordinates for large $R$. The combined limit $\alpha\rightarrow\alpha_\pm,w\rightarrow w_\pm$ is now simply $R\rightarrow0$.

Changing to these new coordinates and expanding around $R=0$ yields the 8d metric
\begin{align}
\dd s^2_8=\frac{1}{2}\bigg[\dd R^2 +R^2\,\bigg(\dd\zeta^2+\cos^2\zeta\,\Big( \big(\dd \psi +\sigma +\tfrac{1}{3}A\big)^2 + \dd s^2_{\text{KE}_4} \Big)+\sin^2\zeta \:\frac{f'(w_\pm)^2}{4(1-j^2)\,w_\pm^2}\, D {z}^2\bigg)\bigg]\,.
\end{align}
Here the gauge field, which now appears as a fibration term in the one-form $\hat{\eta}$ is given by $A=h(w_\pm) \,(\dd z+jr\,\dd t)$. The $\dd z$ leg is constant so it can be absorbed by a shift in $\psi$. Furthermore, the period of $z$ at the poles of the spindle is given by
\begin{equation}
\Delta z = \frac{4\pi\sqrt{1-j^2}\, |w_\pm|}{n_\pm|f'(w_\pm)|} \,.
\end{equation}
If we rescale $z$ such that it is $2\pi$ periodic, i.e. use the coordinate $\hat{z}$ introduced in \eqref{eq:zhat}, we find the metric 
\begin{align}
\dd s^2_8=\frac{1}{2}\bigg[\dd R^2 +R^2\,\bigg(\dd\zeta^2+\cos^2\zeta\,\Big( \big(\dd \hat{\psi} +\sigma+ \tfrac{1}{3}jr\,h(w_\pm)\,\dd t \big)^2 + \dd s^2_{\text{KE}_4} \Big)+\frac{1}{n_{\pm}^2}\sin^2\zeta\, D \hat{z}^2\bigg)\bigg]\,,\label{eq:D2monopole}
\end{align}
where $\hat{\psi}$ and $\hat{z}$ are the shifted and rescaled coordinates with correspondingly rescaled one-form $D\hat{z}$. Note that when the K\"ahler--Einstein space is $\mathbb{CP}^2$ this is $\mathbb{R}^8/\mathbb{Z}_{n_{\pm}}$. We recognize this geometry as a monopole of weight $n_{\pm}$. The presence of these monopoles at the corners of the $(w,\alpha)$ rectangle explains the singularities there and these singularities cannot be resolved. This is most reminiscent of the case of M5-branes on a spindle or disc where the uplifted theory still admits a conical singularity \cite{Ferrero:2021wvk,Bah:2021hei,Bah:2021mzw}. In appendix \ref{app:M5} we have analyzed the M5-brane solution studied there in this light and found the same type of singularity structure.

We emphasise that these monopoles and their corresponding singularities have not been found in the uplift of this 4d setup to 11d supergravity, see \cite{Ferrero:2020twa,Couzens:2021rlk,Couzens:2021cpk,Ferrero:2021ovq}. Instead, in the M2-brane setup, the singularities of the 4d theory are removed in the uplifted theory (under suitable conditions on the fluxes, which are the same ones we take here). Here we find the novel feature that the singularities cannot be removed in a different uplift of the same black hole solution. Moreover, whereas the M5-brane solution realises supersymmetry via a twist, here supersymmetry is realised via an anti-twist and therefore removal of singularities is dependent on the uplift and not the particular twist used.

\subsection{Flux quantization}

We now want to quantize the fluxes of the solution. This is most simply studied in string frame as it removes the need to take into account powers of the dilaton. Consider first the Romans mass, it must satisfy 
\be\label{quantF0}
F_0=\frac{n}{2\pi \ls}\, ,\qquad n\in \mathbb{Z}\, ,
\ee 
in order for the theory to be well-defined. Here $n$ can be interpreted as the number of D8-branes present in the setup.

In order to work out the quantization of the remaining fluxes, we first compute the corresponding Page fluxes. Since the Maxwell charges are generically not-closed they do not give rise to conserved quantities. One should instead consider Page charges since these are closed and thus conserved. The cost of using the Page charges is that they are not gauge invariant quantities, while the Maxwell charges are. The Page fluxes, using polyform notation, are defined as
\be
\hat{f}=F\wedge\me^{-B_2}\, ,
\ee
where $F$ is understood to be the magnetic part of the flux\footnote{Note that in this notation some of these fluxes $F_p$ obtain a minus sign with respect to the Hodge star of the corresponding $(10-p)$ form. This is the case if $p/2$ is odd.}
\be
F=F_0+F_2+F_4+F_6+F_8+F_{10}\Big|_{\text{magnetic}}\, ,
\ee
in string frame. The quantization conditions on the Page fluxes $\hat{f}_p$ read
\be
\frac{1}{(2\pi \ls)^{p-1}}\int_{\Sigma_{p}}\hat{f}_{p}\,\in\, \mathbb{Z}
\ee
where $\Sigma_p$ is an integral cycle in the compact 8d internal space. Let us consider all possible cycles that we can construct in the geometry. There is a single two-cycle which is simply the spindle at a fixed point on the squashed sine cone $Y_6$ which we can take to be at either of the endpoints of the $\alpha$ interval. There are no topological four-cycles in the geometry that we can integrate the four-form flux over. There is a single six-cycle which is given by $Y_6$ at a fixed point on the spindle and a single eight-cycle given by the full internal space.

To proceed we need to choose a gauge for the $B$-field. We take
\be
B_2=F_2 / F_0 \,,
\ee
which can be recognized as a proper choice for $B_2$ directly from the Bianchi identities of massive type IIA, and can also be checked from \eqref{eq:H}-\eqref{eq:F2}. Using this gauge choice we can compute the Page fluxes, e.g. as
\begin{equation}
\hat{f}_6 = -\star F_4 - F_4\wedge B_2 + F_2\wedge \frac{B_2^{{\,2}}}{2} - F_0\,\frac{B_2^{\,3}}{6} \,,
\end{equation}
with $\star$ to be taken in the string frame metric. The quantization of this six-form flux yields the condition
\begin{equation}
\frac{1}{(2\pi \ls)^{5}}\int_{Y_6}\hat{f}_{6} = \frac{1}{(2\pi \ls)^{5}}\,\frac{16L^5}{3m^{1/5}}\,\text{Vol}_{\text{SE}_5} = N \,\in\, \mathbb{Z} \,.
\end{equation}
Here this $N$ can be interpreted as the number of D2-branes which give rise to the 3d theory before being wrapped on the spindle. Together with the condition \eqref{quantF0} on $F_0$, this leads to the following quantization conditions on the parameters $L$ and $m$:
\begin{equation}\label{quantiz}
L = \frac{2^{1/6}3^{5/24}\,\pi\ls \,n^{1/24}N^{5/24}}{\text{Vol}_{\text{SE}_5}^{5/24}}\,, \quad\qquad m = \frac{3^{5/24}\,n^{25/24}N^{5/24}}{2^{5/6}\,\text{Vol}_{\text{SE}_5}^{5/24}} \,.
\end{equation}
Note that these expressions agree with \cite{Fluder:2015eoa} for the parent theory, after taking into account the redefinitions $L_{\text{here}} = 2^{5/16}3^{1/2}\,L_{\text{there}}$ and $m_{\text{here}} = 2^{5/16}\,(\me^{-5\Phi_0/4})_{\text{there}}$. It turns out that the two-form Page flux vanishes. The eight-form flux $\hat{f}_8$ does not vanish, but its integral over the internal space does. The quantization conditions for these fluxes are therefore trivially satisfied and we only have these two flux quanta to consider.

We may now compute the holographic free energy, which is simply given by 
\be
\mathcal{F}_{1d}=\frac{1}{4 G_2}
\ee
with $G_2$ the 2d Newton constant, which is computed by integrating over the volume of the internal space giving
\begin{equation}
\frac{1}{G_2} = \frac{1}{G_{10}}\int_{\mathcal{M}_8}\text{dvol}_{8} = \frac{L^8\,\text{Vol}_{\text{SE}_5}}{3^{3/2}\,5\, \pi^6\ell_s^8}\,\Delta z\,(w_+-w_-) \,,
\end{equation}
where we have used that $G_{10} = 2^3 \pi^6\ell_s^8$. Using the quantization of $L$ \eqref{quantiz}, we find 
\begin{equation}
\frac{1}{G_2} = \frac{2^{4/3}3^{1/6}\pi^2}{5\,\text{Vol}_{\text{SE}_5}^{2/3}}\,\Delta z \,(w_+-w_-)\,n^{1/3}\,N^{5/3} \,.
\end{equation}
The free-energy in terms of the orbifold weights and electric charge is then
\be
\mathcal{F}_{1d}=\frac{  3^{1/6} \pi^3}{2^{2/3} 5 \text{Vol}_{\text{SE}_5}^{2/3}} n^{1/3} N^{5/3} \frac{\sqrt{2(n_-^2+n_+^2) +n_-^2 n_+^2 Q_e^2}-(n_-+n_-)}{n_- n_+}\, .
\ee
This may be further simplified by writing it in terms of the free-energy of the 3d parent theory. Following the conventions in \cite{Fluder:2015eoa} the free energy of the 3d parent theory is
\be
\mathcal{F}_{3d}= \frac{2^{1/3} 3^{1/6}\pi^3}{5 \text{Vol}_{\text{SE}_5}}n^{1/3} N^{5/3}\, ,
\ee
and therefore
\be
\mathcal{F}_{1d}=\frac{\sqrt{2(n_-^2+n_+^2) +n_-^2 n_+^2 Q_e^2}-(n_-+n_-)}{4n_- n_+}\mathcal{F}_{3d}\, .
\ee
It would be interesting to understand this from a dual field theory computation.

\section{M5-branes on SLAG-3 $\times$ spindle}\label{sec:SLAG}

\subsection{The holographic duals of M5-branes on a SLAG-3 manifold}

In the previous section we have studied the uplift of 4d Einstein--Maxwell to massive type IIA using the known uplift formula. There is a further uplift that we will consider which arises from considering M5-branes on a SLAG 3-manifold \cite{Gauntlett:2007ma}. The AdS$_4$ vacuum solution has purely magnetic four-form flux. This may be further generalised by considering the uplift in \cite{Larios:2019lxq}, which uplifts to the most general $\mathcal{N}=2$ AdS$_4$ solutions in 11d supergravity with mixed magnetic and electric fluxes, as classified in \cite{Gabella:2012rc}. We will consider only the uplift to the SLAG 3-manifold theories in the following. We will present the general uplift of the solutions before reducing to considering the 3-manifold to be (a quotient of) $\mathbb{H}^3$.

The metric of the seed 11d AdS$_4$ solutions takes the form
\begin{align}
\dd s^2_{11}=&\frac{1}{m \lambda}\dd s^2(\text{AdS}_4) + \dd s^2(\mathcal{N}_{7})\, ,\nonumber\\
\dd s^2(\mathcal{N}_{7})=& \dd s^2(\mathcal{M}_{\text{SU}(2)}) +\hat{w}^2 +\frac{\lambda^2}{4m^2}\Big(\frac{1}{1-\lambda^3\rho^2}\dd\rho^2 +\rho^2\dd\phi^2\Big)\, ,\label{eq:M5slaggenmetric}
\end{align} 
where $\mathcal{M}_{\text{SU}(2)}$ is a four-dimensional space equipped with a set of SU$(2)$ structure forms which we denote by $J^{a}$ and $\hat{w}$ is a one-form. The metric is supported by the four-form flux
\be
G= \frac{1}{2m}\dd \phi\wedge \dd\bigg[\frac{\sqrt{1-\lambda^3\rho^2}}{\sqrt{\lambda}}J^3\bigg]\, .
\ee
The solution is determined by a solution to the three differential conditions
\begin{align}
\dd \Big[\frac{\sqrt{1-\lambda^3 \rho^2}}{\lambda} \hat{w} \Big]=& \,m\lambda^{-1/2} J^1 +\frac{\lambda^2\rho}{2  \sqrt{1-\lambda^3\rho^2}}\hat{w} \wedge \dd \rho\, ,\nonumber\\
\dd \Big[ \lambda^{-3/2}J^3 \wedge \hat{w} -\frac{\lambda\rho}{2m\sqrt{1-\lambda^3\rho^2}}J^2\wedge \dd \rho\Big]=&\,0\, ,\label{eq:M5eqstosolve}\\
\dd\Big[ J^2\wedge \hat{w}+\frac{\lambda^{-1/2}}{2m\rho \sqrt{1-\lambda^3\rho^2}}J^3\wedge \dd\rho\Big]=&\, 0\nonumber\, .
\end{align}

The KK-reduction of the solution to 4d $\mathcal{N}=2$ Einstein--Maxwell supergravity was performed in \cite{Gauntlett:2007ma}. The truncation ansatz is
\begin{align}
\dd s^2_{11}=&\, \frac{1}{m \lambda}\dd s^2_{4}+\dd s^2(\mathcal{N}^{g}_7)\, ,\nonumber\\
\dd s^2(\mathcal{N}_{7})=&\,\dd s^2(\mathcal{M}_{\text{SU}(2)}) +\hat{w}^2  +\frac{\lambda^2}{4m^2}\Big(\frac{1}{1-\lambda^3\rho^2}\dd\rho^2 +\rho^2\big(\dd\phi+A)^2\Big)\, ,\label{eq:SLAGtrunc}\\
\hat{G}=&\,G^{g}+F\wedge Y +\star_4 F\wedge X\, ,\nonumber
\end{align}
where the subscript `$g$' implies the gauging of the Killing vector $\partial_{\phi}$ by the graviphoton of the 4d theory. At the level of forms this is equivalent to the replacement $\dd\phi\rightarrow \dd\phi+A$. The two two-forms $X,Y$ are given by
\begin{align}
X=&\, -\frac{1}{2m} \Big(\me^{-\tfrac{1}{2}\lambda}J^1 +\frac{\me^{2\lambda}\rho}{2\sqrt{1-\me^{3\lambda} \rho^2}}\hat{w}\wedge \dd\rho\Big)\, ,\label{eq:SLAGtruncforms}\\
Y=&\, -\frac{1}{2m}\me^{-\tfrac{1}{2}\lambda}\sqrt{1-\me^{3\lambda}\rho^2}J^3\, .\nonumber
\end{align}

With the uplift formulae in hand, given a solution to \eqref{eq:M5eqstosolve} it is a simple manner to study the uplifted solution. We will consider the uplift to 11d of a solution originally constructed by Pernici and Sezgin in \cite{Pernici:1984nw} which was subsequently rediscovered in \cite{Gauntlett:2000ng,Acharya:2000mu} by studying wrapped M5-branes. 
The solution is the gravity dual of a stack of M5-branes wrapped on a quotient of 3d hyperbolic space $\mathbb{H}^3$. Since the solution we will study was originally found by studying (a truncation) of maximal 7d gauged supergravity, after the uplift we are able to interpret the solution as the holographic dual of M5-branes wrapped on the 5d space consisting of the direct product of a spindle and $\mathbb{H}^3$.

\subsection{The Pernici--Sezgin solution and uplift}

Let us first consider maximal 7d gauged supergravity. We follow the conventions in \cite{Gauntlett:2000ng}, where the solution we are interested in was rediscovered. The Lagrangian for the bosonic fields is
\begin{align}
2\mathcal{L}=&\, \bigg[ R +\frac{1}{2}m^2 (T^2 -2 T_{ij}T^{ij})-P_{\mu ij}P^{\mu ij}-\frac{1}{2} (\tensor{\Pi}{_{A}^{i}}\tensor{\Pi}{_{B}^{j}}F_{\mu\nu}^{AB})^2 -m^2 \big(\tensor{\Pi}{^{-1}_{i}^{A}} S_{\mu\nu\rho,A}\big)^2\bigg]\dd\vol_7\nonumber\\
&-6 m \delta^{AB} S_A\wedge F_B +\sqrt{3} \epsilon_{ABCDE}\delta^{AG}S_G\wedge F^{BC}\wedge F^{DE}+\frac{1}{8m} (2 \Omega_5[A]-\Omega_3[A])\, ,
\end{align}
where the indices $A,B,..=1,..,5$ denote the indices of the SO$(5)$ gauge group, while latin indices $i,j=1,..,5$ are for the SO$(5)_c$ local composite gauge group and are raised with the Kronecker delta. There are 14 scalars $\tensor{\Pi}{_{A}^{i}}$ living in the coset SL$(5,\mathbb{R})/$SO$(5)_c$ and transform in the $\boldsymbol{5}$ of SO$(5)$ from the left and SO$(5)_c$ from the right. The scalar kinetic term, $P_{\mu ij}$ is the symmetric part of
\be
\tensor{\Pi}{^{-1}_{i}^{A}}\big(\delta_{A}^{B}\partial_{\mu}+ g \tensor{A}{_{\mu}_{A}^{B}} \big)\tensor{\Pi}{_{B}^{k}}\delta_{kj}\, ,
\ee
while the composite gauge field $Q_{\mu i j}$ is the anti-symmetric part. The gauge fields are denoted by $\tensor{A}{_{A}^{B}}$ with field strength $\tensor{F}{_{A}^{B}}$ and coupling constant $g=2m$. The potential terms are 
\be
T_{ij}=\tensor{\Pi}{^{-1}_{i}^{A}}\tensor{\Pi}{^{-1}_{j}^{B}}\delta_{AB}\, ,\qquad T=\delta^{ij}T_{ij}\, .
\ee
The last terms are Chern--Simons terms for the gauge fields, however they will not play a role and so we suppress their details. 

The solutions of interest are of the form AdS$_4\times \mathbb{H}^3$. The $SO(3)$ spin connection of $\mathbb{H}^3$ is identified with an SO$(3)$ subgroup of the R-symmetry via the splitting
\be
\text{SO}(5)\rightarrow \text{SO}(3)\times \text{SO}(2)\, ,\label{eq:SO5splitting}
\ee
with both the scalars and gauge vectors obeying the symmetry. 
The solutions here will have only a single scalar excited which allows us to write the scalars as
\be
\tensor{\Pi}{_{A}^{i}}=\text{diag}\big(\me^{2\Phi},\me^{2\Phi},\me^{2\Phi},\me^{-3\Phi},\me^{-3\Phi}\big)\, .
\ee
The gauge fields take a similar form, becoming only SO$(3)$ valued, i.e. only gauge fields with the indices $1,2,3$ are turned on. Let us split the SO$(5)$ indices as $A=\{a,\hat{a}\}$ with $a=1,2,3$ and $\hat{a}=4,5$, then the gauge fields are
\be
g A_{ab}=\omega_{ab}\, ,
\ee
where $\omega_{ab}$ is the spin-connection on $\mathbb{H}^3$. The metric is
\be
\dd s^2_{7}=\frac{\me^{8 \Phi}}{m^2}\Big(\dd s^2(\text{AdS}_4)+\frac{1}{2}\dd s^2(\mathbb{H}^3)\Big)\, ,
\ee
where the metric on $\mathbb{H}^3$ is normalised to have constant curvature $R_{\mu\nu}=-g_{\mu\nu}$ and the scalar takes the constant value
\be
\me^{10\Phi}=2\, .
\ee

This may be uplifted to 11d supergravity on an $S^4$ using the consistent truncation in \cite{Pernici:1984xx,Nastase:1999cb}. The metric is given by
\be
\dd s^2_{11}=\Delta^{-2/5} \dd  s_{7}^2+\frac{1}{m^2}\Delta^{4/5}\Big[\me^{4\Phi} DY^a D Y^a+\me^{-6 \Phi}\dd Y^{\hat{a}}\dd Y^{\hat{a}}\Big]\,,\label{eq:11duplift}
\ee
where the $Y^{A}$ are embedding coordinates for the $S^4$ satisfying $Y^{a}Y^{a}+Y^{\hat{a}}Y^{\hat{a}}=1$. The gauged one-forms are defined to be
\be
D Y^{a}=\dd Y^a +2m A^{ab} Y^b\, ,
\ee
and the warp factor is
\be
\Delta^{-6/5}= \me^{-4\Phi}Y^aY^a+\me^{6\Phi}Y^{\hat{a}}Y^{\hat{a}}\, .
\ee
The expression for the flux will be given momentarily. 

To make contact with the form of the metric in \eqref{eq:M5slaggenmetric} we should introduce explicit coordinates for the embedding coordinates $Y^A$. Following \cite{Gauntlett:2006ux} we introduce the coordinates
\be
Y^{a}=\sqrt{1-\frac{\rho^2}{8}}\mu^a\, ,\qquad Y^4=\frac{1}{2\sqrt{2}}\rho \sin\phi\, ,\qquad Y^5=\frac{1}{2\sqrt{2}}\rho\cos\phi\, ,
\ee
with $\mu^a$ embedding coordinates for an $S^2$. 
The scalar $\Delta$ takes the form
\be
\Delta^{-6/5}=\frac{\me^{-4\Phi}}{8}(8+\rho^2)\, .
\ee
The final metric is then
\begin{align}
\dd s^2_{11}=\me^{-\lambda}\bigg[\dd s^2(\text{AdS}_4)+\frac{1}{2}\dd s^2(\mathbb{H}^3)+\frac{8-\rho^2}{8+\rho^2}D \mu^aD\mu^a+\frac{1}{2}\Big(\frac{\rho^2}{8+\rho^2}\dd\phi^2 +\frac{1}{8-\rho^2}\dd\rho^2\Big)\bigg]\, ,
\end{align}
while the flux is given by
\be
G=\frac{1}{2}\dd\phi\wedge \dd \bigg[\me^{-\tfrac{1}{2}\lambda}\sqrt{1-\me^{3\lambda}\rho^2}J_3\bigg]\, ,
\ee
where we defined
\be
\me^{3\lambda}=\frac{2}{8+\rho^2}\, ,
\ee
and the two-form $J_3$ takes the form
\be
J_3=\frac{\me^{-\lambda}}{8}\epsilon^{abc}\mu^{a}\Big(\big(1-\me^{3\lambda}\rho^2)D\mu^b\wedge D\mu^c- \frac{1}{2} e_{\mathbb{H}^3}^{b}\wedge e_{\mathbb{H}^3}^{c}\Big)\, .
\ee
For later it is useful to note that the quantisation of the flux through the $S^4$ gives the number of M5-brane
\be
\frac{1}{(2\pi\lp)^3}\int_{S^4}G=\frac{1}{\pi m^3 \lp^3}\equiv N\in \mathbb{Z}\, ,
\ee
and that the free energy is 
\be
\mathcal{F}_3=\frac{N^3}{3\pi}\text{Vol}(\mathbb{H}^3)\, .
\ee

Having given the AdS$_4$ solution, and noted that the R-symmetry vector is written explicitly, we can now truncate to 4d Einstein--Maxwell supergravity using the consistent truncation given in \cite{Gauntlett:2007ma} and reviewed in the previous section. We may further view this as a truncation of maximal 7d gauged supergravity on $\mathbb{H}^3$ to 4d Einstein--Maxwell with a little rewriting. The presence of the 4d graviphoton leads to a gauging of the embedding coordinates $Y^{\hat{a}}$ in \eqref{eq:11duplift}. That is we turn on a new gauging
\be
\dd Y^{\hat{a}}\rightarrow DY^{\hat{a}}=\dd Y^{\hat{a}}+ A^{\hat{a}\hat{b}}Y^{\hat{b}}\, ,
\ee 
where the only non-trivial component is
\be
A^{45}=A\,,
\ee
with $A$ the graviphoton of the 4d theory. 
From the 7d truncation this amounts to keeping the same scalars of the seed theory and modifying the gauge fields to include
an extra component in the SO$(2)$ of the SO$(5)$ splitting we performed in \eqref{eq:SO5splitting}. where $A$ is the graviphoton of the 4d theory. By construction, and the results of \cite{Gauntlett:2007ma} this leads to a consistent truncation of the 7d theory on $\mathbb{H}^3$ to 4d Einstein--Maxwell supergravity.

\subsection{AdS$_2\ltimes \Sigma\times \mathbb{H}^3$ solutions}

We may now use the consistent truncation outlined above of 7d maximal gauged supergravity to 4d Einstein--Maxwell supergravity to uplift the spindle solution of section \ref{eq:EMspindle}. 
The 7d solution is then
\begin{align}
\dd s^2=&\,\me^{8 \Phi}\Big(\sqrt{P}(w) \Big(\dd s^2(\text{AdS}_2)+\dd s^2_{\Sigma}\Big)+\frac{1}{2}\dd s^2(\mathbb{H}^3)\Big)\, ,\\
A^{AB}=&\,\begin{pmatrix}
 \omega^{ab}& 0\\
 0 & \mathcal{A}^{\hat{a}\hat{b}}
 \end{pmatrix}\, ,\\
\tensor{\Pi}{_{A}^{i}}=&\,\text{diag}\big(\me^{2\Phi},\me^{2\Phi},\me^{2\Phi},\me^{-3\Phi},\me^{-3\Phi}\big)\, ,
\end{align}
where
\be
\mathcal{A}^{\hat{a}\hat{b}}=\epsilon^{\hat{a}\hat{b}}h(w) (\dd z+j r \dd t)\, ,\qquad \me^{10\Phi}=2\, .
\ee

Uplifting to 11d supergravity we find the (fibered) AdS$_2$ metric\footnote{In \cite{Couzens:2020jgx} fibered AdS$_2$ solutions of 11d supergravity with electric four-form flux were classified. The uplift of the spindle solution we consider here to 11d supergravity on a 7d Sasaki--Einstein manifold as studied in \cite{Ferrero:2020twa} can be embedded into the classification. However since we have magnetic flux here these solutions are outside the classification in \cite{Couzens:2020jgx}. It would be interesting to extend the classification allowing for all fluxes turned on, giving a fibered AdS$_2$ version of \cite{Hong:2019wyi}, in the same spirit that  \cite{Couzens:2020jgx} gave a fibered version of \cite{Kim:2006qu} }
\begin{align}
\dd s^2_{11}=&\, \me^{-\lambda}\bigg[\sqrt{P(w)}\Big(\dd s^2(\text{AdS}_2)+\dd s^2(\Sigma)\Big)+\frac{1}{2}(\mathbb{H}^3)+\frac{8-\rho^2}{8+\rho^2}D\mu^aD\mu^a \nonumber\\
&\quad+\frac{1}{2}\Big(\frac{\rho^2}{8+\rho^2}\big(\dd \phi+h(w)(\dd z+j r \dd t)\big)^2+\frac{1}{8-\rho^2}\dd\rho^2\Big)\bigg]\, .
\end{align}
The can be read off from \eqref{eq:SLAGtrunc} and \eqref{eq:SLAGtruncforms} where the forms appearing in \eqref{eq:SLAGtruncforms} are 
\begin{align}
J^1=&\frac{\sqrt{1-\me^{3\lambda}\rho^2}}{2^5/2\me^{\lambda}}D\mu^a\wedge e_{\mathbb{H}^3}^a\, ,\qquad
\hat{w}=\frac{\me^{-\tfrac{1}{2}\lambda}}{2^{3/2}}\mu^a e^{a}_{\mathbb{H}^3}\, .
\end{align}

Let us now analyse the regularity of the solution. Since we performed the regularity analysis of the D2-brane solutions in great detail we will refer to some of the results there and keep the regularity analysis lighter here. We will restrict to studying the metric at a fixed point in AdS$_2$ and ignore the regularity conditions involving $\mathbb{H}^3$ since this is well established in the literature. 

\begin{figure}[h!]
\centering
  \includegraphics[width=0.75\linewidth]{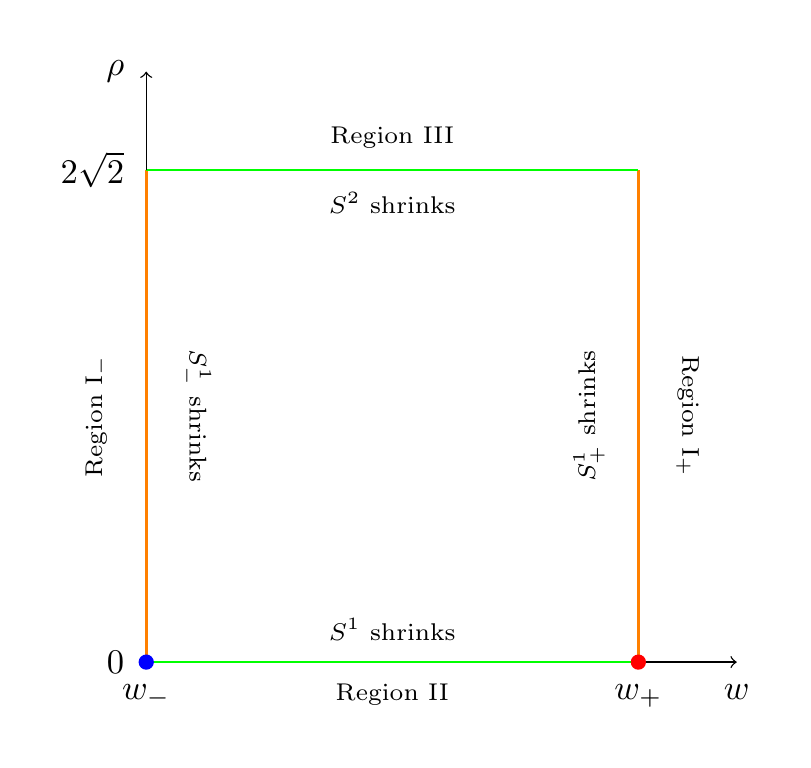}
  \captionsetup{width=.85\linewidth}
  \caption{\textit{A schematic plot of the rectangle over which there is an $S^1\times S^1\times S^2$ fibration.  There are three  distinguished regions, plus two special points (the dots at the bottom corners) giving rise to conical defects/monopoles. The red monopole has weight $n_{+}$ whilst the blue monopole has weight $n_-$.}}
  \label{fig:M5rectangle}
\end{figure}

We may break the regularity analysis into three regions on the boundary of a rectangle as depicted in figure \ref{fig:M5rectangle}. The range of the coordinate $\rho$ is $\rho\in[0,2\sqrt{2}]$, and $\rho=0$ defines region II, while its maximal value at at $2\sqrt{2}$ defines region III. The regions I$_{\pm}$ are located at the north and south poles of the spindle. 

\paragraph{Region I}

Let us consider a pole of the spindle. Similar to section \ref{sec:D2reg} we may construct the following suitably normalised Killing vectors with vanishing norm at one of the poles of the spindle:
\be
\partial_{\phi_{\pm}}=\frac{n_{\pm}\Delta z }{2\pi}\Big(\partial_z-h(w_{\pm})\partial_{\phi}\Big)\, .
\ee
Away from $\rho=0$ where $\phi$ shrinks this gives rise to a smooth $S^3$. To see this we should write the metric in a similar form to the one presented in \eqref{eq:D2S3}. The change of coordinates is
\begin{align} 
z=\frac{n_+\Delta z}{2\pi}\phi_++\frac{n_-\Delta z}{2\pi}\phi_-\, ,\quad \phi=-\frac{n_+h(w_+)\Delta z}{2\pi}\phi_+-\frac{n_- h(w_-)\Delta z}{2\pi}\phi_-\, . 
\end{align}
Then the spindle and circle part of the metric becomes
\begin{align}
&\sqrt{P}(w)\dd s^2(\Sigma)+\frac{\rho^2}{2(8+\rho^2)}\big(\dd \phi+h(w)(\dd z+j r \dd t)\big)^2= \nonumber\\
&\sqrt{P(w)}{f(w)}\dd w^2+F_+(w,\rho)D\phi_+^2+F_-(w,\rho)D\phi_-^2+GD\phi_-D\phi_+
\end{align}
where
\begin{align}
F_{\pm}(w,\rho)&=\frac{n_\pm^2\Delta z^2 \big( 2(8+\rho^2)f(w)+\rho^2 (h(w)-h(w_\pm))^2 \sqrt{P(w)}\big)}{8 \pi^2 (8+\rho^2)\sqrt{P(w)}}\, ,\\
D\phi_{\pm}&=\dd\phi_{\pm}-u_{\pm}j r \dd t\, ,\\
u_\pm&=\pm\frac{2 \pi h(w)}{n_\pm \Delta z (h(w_+)-h(w_-)}\, ,\\
G&=\frac{n_- n_+\Delta z^2\big( \rho^2 \big(h(w)-h(w_-)\big)\big(h(w)-h(w_+)\big)\sqrt{P(w)}+2 f(w)(8+\rho^2)\big)}{4\pi^2 (8+\rho^2)\sqrt{P(w)}}\, .
\end{align}
Note that $F_{\pm}(w_{\pm},\rho)=0$ and $G(w_{\pm})=0$. This then exhibits the metric as a smooth $S^3$ as can be seen by going to the points $w=w_{\pm}$. Note that this is only true away from $\rho=0$.

\paragraph{Region II}

At $\rho=0$ the circle parametrised by $\phi$ shrinks smoothly if it has period $2\pi$. This is of course the expected period coming from the embedding coordinates we used above. At the corners of the rectangle in figure \ref{fig:M5rectangle} there are conical singularities of the same type encountered in the previous section. One can see this by taking the double limit $w=w_{\pm}$ and $\rho=0$, however since this proceeds in a similar fashion to the D2 brane case we will suppress the computational details. One can understand this conceptually by noting that the smoothing out of the spindle singularities requires another circle so that the space becomes topologically an $S^3$. If this circle shrinks somewhere the desingularisation through fibration is not possible at this point, one is left with a copy of the spindle and thus orbifold singularities.

\paragraph{Region III}

 At $\rho=2\sqrt{2}$ the $S^2$ shrinks smoothly. Note that this point along the boundary is perfectly smooth, including the corners, it does not have conical deficits there. This is because the circle fibration over the spindle is left of finite size on this boundary. This should be contrasted with the D2 brane of the previous section where the singularities are present at all four corners.

\paragraph{Flux quantisation}

Let us now consider the quantisation of the four-form flux. The only four-cycle is obtained from the $S^4$ at fixed point on the spindle and $\mathbb{H}^3$. It is simple to show that
\be
\frac{1}{(2\pi\lp)^3}\int_{S^4} G=\frac{1}{\pi m^3 \lp^3}\equiv N\in\mathbb{Z}\, ,
\ee
where $N$ should should be understood as the number of M5-branes being wrapped.

\paragraph{Free Energy}

Finally let us compute the free energy of the solution. As before the free energy can be extracted from evaluating the 2d Netwon's constant, which is given by
\begin{align}
\mathcal{F}_{1}=\frac{1}{4G_2}=&\frac{1}{4G_{10}}\int_{\mathcal{M}_9}\dd\vol_9=\frac{\Delta z (w_+-w_-)}{4}\mathcal{F}_{3}\nonumber\\
&= \frac{\sqrt{2(n_-^2+n_+^2) +n_-^2 n_+^2 Q_e^2}-(n_-+n_-)}{4n_- n_+}\mathcal{F}_3\, .
\end{align}

\section{Concluding remarks}\label{Sec:Conclude}

In this paper we have considered the holographic duals of a large class of 3d SCFTs wrapped on a spindle. We have exploited the consistent truncations of massive type IIA and 11d supergravity on a SLAG manifold to 4d $\mathcal{N}=2$ Einstein--Maxwell supergravity and uplifted the known rotating Plebanski--Demianski metric \cite{Plebanski:1976gy}. We find that whereas the uplift of the same solution to 11d supergravity on an $S^7$ allows for the conical singularitites of the spindle to be resolved through fibration, this is not possible in either of the two geometries studied here. Rather, the in the massive type IIA case the 10d metric has four singularities consistent with the presence of KK-monopoles and is similar to the singularities arising from M5-branes on a spindle. On the other hand the solutions holographic dual to M5-branes wrapped on a spindle and $\mathbb{H}^3$ have only two orbifold singularities, rather than the four of the D2-brane case. 
One can understand this by considering the degenerations of the internal manifold on which the solutions have been uplifted. For odd dimensional spheres there is always a circle which remains finite, over which we may fibre the spindle to obtain a smooth $S^3$. For even dimensional spheres, there are points where all circles shrink, and it is at these points that we are left with a copy of the spindle and therefore orbifold singularities. 
Our second solution has extended the known examples of wrapped branes on products of spindles with constant curvature spaces, see \cite{Suh:2022olh,Boido:2021szx} for the direct product of spindle and a Riemann surface. It would be interesting to construct solutions where the 5d space on which the M5-brane is wrapped is a full spindle, in the same spirit as the 4d orbifold solutions of \cite{Cheung:2022ilc}.

It is natural to interpret the conical defects of the spindle as regular punctures of a round two-sphere. Indeed, in the case of M5-branes on a disc this has been studied in \cite{Bah:2021hei,Bah:2021mzw,Couzens:2022yjl} and matched to a dual field theory computation. The disc geometry is understood to be a two-sphere with a regular puncture at one pole and an irregular puncture at the other. Due to the presence of the irregular puncture the R-symmetry of the field theory must mix with the isometries of the two-sphere, which is a well established fact on the gravity side for spindle geometries. 
The disc solutions considered in \cite{Bah:2021hei,Bah:2021mzw} can be obtained as a special limit of the local solution giving rise to the holographic dual of M5-branes on a spindle \cite{Ferrero:2021wvk}, see for example \cite{Couzens:2022yjl}. In \cite{Bah:2021hei,Bah:2021mzw}  it was shown that the conical defect gives rise, on the field theory side, to a flavour symmetry described by a rectangular Young diagram. This was further extended in \cite{Couzens:2022yjl} to allow for an arbitrary regular puncture, therefore generalising the conical defect of the disc. One should expect that a similar type of generalisation is possible for spindle solutions too.


\section*{Acknowledgments}

We would like to thank Damian van de Heisteeg for discussions and previous collaboration on related topics.
CC would also like to thank Hyojoong Kim, Nakwoo Kim, Yein Lee and Minwoo Suh for discussions and collaboration on related work, past and present. CC is supported by the National Research Foundation of Korea (NRF) grants 2019R1A2C2004880, 2020R1A2C1008497.

\appendix


\section{M5-branes on a spindle}\label{app:M5}

In the main text we have claimed that the singularity structure of the D2-brane on a spindle solutions constructed here are more reminiscent of the solutions for M5-branes on a spindle. To aid comparison we will review the AdS$_5$ solutions of 11d supergravity corresponding to M5-branes on a spindle, as first studied in \cite{Ferrero:2021wvk}. Our interest in this solution is to study the singular behaviour of the uplifted theory and to compare with our analysis of the solutions corresponding to D2-branes on a spindle. 

The 7d metric of \cite{Ferrero:2021wvk} is
\begin{align}
\dd s^2&=\Big(w P(w)\Big)^{1/5}\bigg[4 \dd s^2(\text{AdS}_5)+\frac{w}{f(w)}\dd w^2+\frac{f(w)}{P(w)}\dd z^2\bigg]\, .\label{eq:7dM5sol}
\end{align}
The metric on AdS$_5$ is taken to be the one with unit radius which implies that it is Einstein satisfying $R_{\mu\nu}=-4 g_{\mu\nu}$. The functions appearing in the metric are the simple polynomials
\begin{align}
h_{i}(w)&=w^2-l_i\, ,\qquad 
P(w)=h_1(w)h_2(w)\, ,\qquad
f(w)=P(w)-w^3\, .\
\end{align}
The solution is supported by two real scalars and two abelian gauge fields
\be
A_{i}=\frac{l_{i}}{h_i(w)}\dd z\, ,\qquad X_{i}(w)=\frac{\Big(w P(w)\Big)^{2/5}}{h_i(w)}\, ,
\ee

As explained in \cite{Ferrero:2021wvk} the $w$ coordinate is bounded between two positive roots, $w_{\pm}$ of the function $f(w)$. The period of the circle direction $z$ is fixed to be
\be
\frac{\Delta z}{2\pi}= \frac{2 w_{\pm}^2}{n_{\pm}|f'(w_{\pm})|}\, ,
\ee
with $n_{\pm}$ orbifold weights for the spindle at the respective poles. The two magnetic charges are
\begin{align}
Q_{i}=\frac{1}{2\pi}\int_\Sigma\dd A_i=\frac{\Delta z}{2\pi}\Big[\frac{l_i}{h_i(w_+)}-\frac{l_i}{h_i(w_-)}\Big]\, .\label{eq:M5mag}
\end{align}

The solution can be uplifted to 11d supergravity on an $S^4$, with resultant 11d metric
\begin{align}
\dd s^2_{11}= &\Omega^{1/3}\Big(w P(w)\Big)^{1/5}\bigg[ 4 \dd s^2 (\text{AdS}_5)+\frac{w}{f(w)}\dd w^2 +\frac{f(w)}{P(w)}\dd z^2\nonumber\\
&+\frac{1}{\Omega \big(w P(w)\big)^{1/5}}\Big(X_0^{-1}\dd \mu_0^2 + \sum_{i=1}^{2} X_i^{-1}\big(\dd \mu_i^2+\mu_i^2 (\dd\phi_i+A_i)^2\big)\Big)\bigg]\, .
\end{align}
Here, $\mu_I$ are embedding coordinates for the $S^4$ and satisfy $\mu_0^2+\mu_1^2+\mu_2^2=1$. The new functions appearing in the metric are
\begin{align}
X_0=X_1^{-2}X_2^{-2}\, ,\qquad \Omega=\sum_{I=0}^{2}X_I \mu_I^2\, .
\end{align}
It is useful to reparametrise the $\mu_I$ as 
\be
\mu_{1}=\sqrt{1-\mu_0^2} \sin\theta \, ,\quad \mu_{2}=\sqrt{1-\mu_0^2} \cos\theta\, .
\ee
With these coordinates the metric takes the form
\begin{align}
\dd s_{11}^2&= \Omega^{1/3}(w P(w))^{1/5}\Bigg[ 4 \dd s^2(\text{AdS}_5)+\frac{w}{f(w)}\dd w^2+\frac{f(w)}{P(w)}\dd z^2\\
&+ \frac{1}{\Omega(wP(w))^{1/5}}\bigg\{ (1-\mu_0)^2 \bigg[\frac{\sin^2\theta}{X_1} D\phi_1^2+\frac{\cos^2\theta}{X_2} D\phi_2^2+ \Big(\frac{\cos^2\theta}{X_1}+\frac{\sin^2\theta}{ X_2} \Big)\dd \theta^2\bigg]\nonumber\\
&+\frac{2 \mu_0 \sin\theta\cos\theta(X_1-X_2)}{X_1 X_2}\dd \theta \dd\mu_0+\bigg[X_1^2 X_2^2+\frac{\mu_0^2}{1-\mu_0^2}\Big(\frac{\sin^2\theta}{X_1}+\frac{\cos^2\theta}{X_2}\Big)\bigg]\dd\mu_0^2\bigg\}\Bigg]\, ,
\end{align}
with 
\be
D\phi_i=\dd\phi_i+A_i\, ,
\ee
and $\Omega$ takes the form
\be
\Omega=\frac{\mu_0^2}{X_1^2 X_2^2}+(1-\mu_0^2)\Big(X_1\sin^2 \theta +X_2\cos^2\theta\Big)\, .
\ee

Near to $\theta=0,\pi$ an $S^2$ in the metric shrinks smoothly when $\phi_i$ have period $2\pi$, this is of course the expected degeneration of a three-sphere. Next consider a degeneration at the root of $f(w)$ at $w_{\pm}$. The degenerating Killing vector with $2\pi$ period at one of the poles of the spindle (away from $\mu_0^2=1$) is
\be
k_{\pm}=\pm\frac{2 w_{\pm}^2}{|f'(w_{\pm})|}\bigg(\partial_z -\sum_{i=1}^{2}\frac{l_i}{h_i(w_{\pm})}\partial_{\phi_i}\bigg)\, .
\ee
Note that the prefactor satisfies
\be
\frac{2 w_{\pm}^2}{|f'(w_{\pm})|}=\frac{n_{\pm}\Delta z}{2\pi}\, .
\ee
We have four Killing vectors which degenerate at $\theta=0,\pi$ and $w=w_{\pm}$ and therefore they must satisfy a linear relation
\be
a_+ k_++a_-k_-+p_1 \partial_{\phi_1}+p_2\partial_{\phi_2}=0\, .
\ee
This implies three constraints on the parameters:
\be
n_+ a_+-n_- a_-=0\, ,\quad n_+ a_+ Q_{i}=p_{i}\, ,
\ee
where we have used the first condition to simplify the second and the explicit form of the magnetic charges in \eqref{eq:M5mag}. 
Clearly we should solve the first condition by taking
\be
a_+=n_-\, ,\quad a_-=n_+\, ,
\ee
then we find the quantisation condition
\be
n_+ n_-Q_{i}=p_i\in \mathbb{Z}\, .
\ee
The degeneration at $\mu_0=\pm1$ leads to the shrinking of a warped $S^3$ as expected for the degeneration of an $S^4$. 
The manifold is thus smooth and singularity free away from the four points $(\mu_0,w)=(\pm1, w_{\pm})$ which we are yet to study. Let us consider these points and for simplicity set $X_1=X_2$.\footnote{For $X_1\neq X_2$ the coordinate transformation is more complicated due to the mixed term between $\mu_0$ and $\theta$ and the coordinate transformation requires a coordinate transformation for the three coordinates rather than just the two. Strictly the spindle does not exist when $X_1=X_2$ since the required number of positive roots is not possible for the polynomial $f(w)$. However since we are interested in the degeneration at one end-point of the spindle, and not the full global regularity, we are justified to consider $X_1=X_2$.} We should change coordinates to
\be
w= w_{\pm}\mp\frac{|f'(w_{\pm})|X_1(w_\pm)}{2 w_{\pm}^{9/5}} r^2 \cos^2\zeta\, ,\qquad \mu_0=\pm 1\mp r^2\sin^2\zeta\, .
\ee
Then
\begin{align}
\dd s^2_6=\frac{2 X_1(w_{\pm})}{w_{\pm}^{4/5}}\bigg[\dd r^2 +r^2\Big\{\dd \zeta^2+\frac{1}{n_{\pm}^2}\cos^2\zeta \dd \hat{z}^2+\sin^2\zeta \Big(\dd\theta^2+\sin^2\theta\dd\phi_1^2+\cos^2\theta\dd\phi_2^2\Big)\Big\}\bigg]\, ,
\end{align}
where we defined the $2\pi$ periodic coordinate $\hat{z}$ as
\be
z=(\Delta z)~\hat{z}\, ,
\ee
and we have performed a linear shift of the $\phi_i$ coordinates in the final result. 
This is the metric on $\mathbb{R}^6/\mathbb{Z}_{n_{\pm}}$ and implies the existence of monopoles located at the four points $(\mu_0,w)=(\pm1, w_{\pm})$, with the orbifold weight correlated to the $\pm$ of $w_{\pm}$. This remains when relaxing $X_1=X_2$ and thus is also true for the spindle solution. Compare this to the D2-brane case studied in the main text, in particular equation \eqref{eq:D2monopole}. Note that both have the same phenomenon of resolving the potential line of singularities in the uplift at the cost of singularities at isolated points instead. This should be contrasted with the M2-brane and D3-brane solutions where the uplifted metric can be made smooth everywhere.





\bibliographystyle{JHEP}

\bibliography{universalbib}

\end{document}